\documentclass[twocolumn,amsmath,amssymb,nofootinbib,superscriptaddress]{revtex4-2}

\usepackage{graphics,amssymb,amsmath,epsfig,color,textgreek}
\usepackage{graphicx}
\usepackage[dvipsnames]{xcolor}
\usepackage{dcolumn}
\usepackage{bm}
\usepackage[colorlinks=true,citecolor=cyan]{hyperref}
\hypersetup{colorlinks=true,citecolor=cyan,linkcolor=red,urlcolor=magenta}
\usepackage{braket}
\usepackage[normalem]{ulem}
\usepackage{cancel}
\usepackage{xfrac} 
\usepackage{amsfonts}
\usepackage{amsmath}
\usepackage{amssymb}
\usepackage{bm}
\usepackage{svg}
\usepackage{lipsum}
\usepackage{chemformula}

\usepackage{enumitem}
\newlist{abbrv}{itemize}{1}
\setlist[abbrv,1]{label=,labelwidth=1.1in,align=parleft,itemsep=0.1\baselineskip,leftmargin=!}

\newcommand{\LK}[1]{\textcolor{black}{#1}}
\newcommand{\ham}{\mathcal{H}}

\newcommand{\ecH}{\mathcal{H}}
\newcommand{\ecS}{\mathcal{C}}
\newcommand{\localS}{S}

\begin{document} 

\title{Quantum Eigenvector Continuation for Chemistry Applications}

\author{Carlos Mejuto-Zaera}
\email{cmejutoz@sissa.it}
\affiliation{Scuola Internazionale Superiore di Studi Avanzati (SISSA), Trieste TS, Italy}

\author{Alexander F. Kemper}
\email{akemper@ncsu.edu}
\affiliation{Department of Physics, North Carolina State University, Raleigh, North Carolina 27695, USA}

\date{\today}

\begin{abstract}
    A typical task for classical and quantum computing in chemistry is finding a
    potential energy surface (PES) along a reaction coordinate, which involves 
    solving the quantum chemistry problem for many points along the reaction 
    path.
    Developing algorithms to accomplish this task on quantum computers has been
    an active area of development, yet finding
    all the relevant eigenstates along the reaction coordinate
    remains a difficult problem, and determining PESs
    is thus a costly proposal.
    In this
    paper, we demonstrate the use of a eigenvector continuation --- a subspace
    expansion that uses a few eigenstates as a basis --- as a tool for rapidly
    exploring potential energy surfaces.  We apply this to determining the binding PES 
    or torsion PES for several molecules of varying complexity.  In all cases, we show
    that the PES can be captured using relatively few basis states; suggesting that
    a significant amount of (quantum) 
    computational effort can be saved by making use of already
    calculated ground states in this manner.
\end{abstract}
\maketitle

\section{Introduction}

A central motive of quantum physics and chemistry is the accurate determination of the low lying energy eigenstates of a Hamiltonian describing a system of interest.  
Whether for finding the ground state
of a highly degenerate spin system, such as in spin liquids, or for studying a pathway for
a particular chemical reaction, one inevitably uses the eigenstates and expectation values computed with them.
Unfortunately, finding the energies and eigenstates is a computationally challenging problem, lying in either
the NP-hard (classical) or QMA (quantum) complexity class~\cite{bharti2022noisy}. A number of classical
and quantum algorithms for finding low-lying eigenstates have been developed~\cite{Helgaker2000,Onida2002,martin2004,Schollwoeck2005, Kotliar2006,Schollwock2011,szabo2012modern,martin2016interacting,becca2017,cao2019,bauer2020,mcardle2020,motta2022emerging},
addressing specific difficulties of the problem, but an approach working for all systems and computational demands is so far non-existing.

A common instance of the Hamiltonian eigenstate problem concerns the situation in which one requires the ground and/or excited states as a function of some
Hamiltonian parameter(s) $\boldsymbol{\lambda}$.  For example, such a parameter could be a reaction coordinate 
in a chemical process, or an interaction strength coefficient when investigating the phase transitions in a spin model in condensed matter physics.
In either case, the quantities of interest lie along some path
through parameter space. Given that finding the eigenstates of a single Hamiltonian is already complex, it is easy to imagine that doing so with consistent accuracy along a Hamiltonian path is a daunting endeavor.
In some reasonably common cases, this problem can be simplified.
In particular, unless there is a symmetry-protected level crossing,
the eigenvectors and eigenvalues along a given Hamiltonian path are continuous~\cite{frame2018eigenvector}.
This allows making use of previously computed eigenstates at a set of parameters $\Lambda = 
\left\{ \boldsymbol{\lambda}_1, \ldots, \boldsymbol{\lambda}_N\right\}$ along the path
as an efficient subspace
basis in which to represent the Hamiltonian at some new $\boldsymbol{\ell} \notin \Lambda$.
As long as $N$ is not exponentially large, this subspace is a much smaller problem to handle
than the full diagonalization at $\boldsymbol{\ell}$, and can thus be performed classically at negligible computational cost. This approach, first introduced
in the context of nuclear physics~\cite{frame2018eigenvector}, is named ``eigenvector continuation'' (EC), and it has been further extended to condensed matter physics~\cite{herbst2022surrogate,brehmer2023reduced}
and quantum computing~\cite{francis2022subspace}.

Since it requires computing eigenstates at a limited number of points,
EC may be particularly fruitful in situations where doing so is
computationally expensive.  Quantum computing may be such a case, as the currently viable algorithms
for finding the ground state --- including the variational quantum eigensolver (VQE), adiabatic state preparation (ASP),
or quantum approximate adiabatic optimization (QAOA) --- are expensive hybrid quantum-classical algorithms
that are difficult to converge.  And yet, to compute e.g. the binding curve of a molecule~\cite{kandala2017hardware,parrish2019quantum,stair2020multireference,Qiskit-Textbook,cohn2021quantum},
or the phase diagram of a quantum magnet~\cite{nishimoto2013controlling,broholm2020quantum}, a new iterative loop is
started at each new parameter point. While the initial guess for the loop may be improved,
no further information is carried forward between iterations.
Given the high cost of each eigenstate calculation, being able to reuse previously obtained
eigenvectors could be a great advantage~\cite{herbst2022surrogate,francis2022subspace,brehmer2023reduced}.
The main strategy is thus to perform the costly, exact eigenstate determination in only a small number of parameter points, and then reconstruct the eigenstates in the full path using a reduced, effective Hamiltonian representation in the basis of these selected points.

EC has been successfully employed to model Hamiltonians for solid state systems \LK{on a quantum computer\cite{francis2022subspace}, but} potential applications in the scope of computational chemistry mostly unexplored. 
In this work we address this issue and demonstrate that EC can be readily applied to computing
the binding curves of a number of chemical compounds, with the eye towards applying this to quantum
computing.
We study singly-bonded (\ch{F2}, \ch{HF}) doubly-bonded (\ch{H2CO}, \ch{O2}) and triply-bonded 
(\ch{N2}, {CO}) molecules, as well as more strongly correlated examples (\ch{C6H8}-torsion, \ch{Cr2}). Within
the context of these molecules, we investigate the use of EC and details of its implementation, particularly the special considerations that are unique to the \emph{ab initio} setting.
\LK{
We evaluate the problems entirely on a classical simulator, however, the
method pertains to quantum computation in the same sense that all
subspace methods do\cite{klymko2021real,shen2022real,cortes2022quantum}. It is relatively expensive to find the ground
state of a model at a single parameter point\cite{tilly2022variational} due to a combination of a
large variational search space\cite{bittel2021training,anschuetz2022quantum} barren plateaus\cite{mcclean2018barren,cerezo2021cost,Larocca2022diagnosing}, and deep circuits that are 
not amenable to today's hardware, etc.  Thus, it is beneficial to reduce
the number of times that this task needs to be performed.  We also
note that EC can be used regardless of the particular circuit
ansatz (such as the Hamiltonian Variational Ansatz\cite{wiersema2020exploring}, unitary coupled cluster theory\cite{anand2022quantum}, or ADAPT-VQE\cite{grimsley2019adaptive})
used for the ground state wavefunction, making it quite a general
method.}

\section{Glossary}

We summarize the main choices of notation and nomenclature used throughout the paper in the following list.

\begin{abbrv}
\item[Point in parameter space $\boldsymbol{\lambda}$, $\boldsymbol{\ell}$] A vector containing all values defining a single point in the parameter space of the system studied. 
In this paper, these correspond to all the relative atomic coordinates describing the molecular geometry. In a spin Hamiltonian, these would correspond to the spin-spin couplings.
\item[Local atomic orbitals (AOs)\\$\left\lbrace\phi\right\rbrace_a$] The basis for the quantum chemistry problem at a particular atomic configuration $\boldsymbol{\lambda}$. These are typically not orthogonal.
\item[Fock operator\\$F_a$] Fock operator in the basis local AOs at a particular $\boldsymbol{\lambda}_a$. Plays the role of the Hamiltonian in the non-linear, generalized eigenvalue problem of the Hartree-Fock approximation.
\item[Local overlap\\$\localS_a$] Overlap integrals of a single local AO basis set at a particular $\boldsymbol{\lambda}_a$.
\item[Local molecular orbitals (MOs)] The (orthogonal) molecular orbitals found by solving the Hartree-Fock generalized eigenvalue equation.
\item[Local orbital rotation matrix $U$] The unitary rotation matrix that diagonalizes the Hartree-Fock Hamiltonian, and rotates from AOs to MOs.
\item[FCI Hamiltonian $H^{FCI}_a$] Full configuration interaction Hamiltonian in the basis of molecular orbitals (MOs) at a particular $\boldsymbol{\lambda}_a$.
\item[FCI rotation matrix $Q$] The unitary rotation matrix that diagonalizes the FCI Hamiltonian.
\item[EC training vector $\ket{v^{(n)}_i}$] $n$-th eigenstate of the FCI Hamiltonian at a particular training point $\boldsymbol{\lambda}_i$.
\item[EC overlap\\$\ecH_{ij}(\boldsymbol{\ell})$] Hamiltonian matrix element evaluated with the training state vectors $\mathcal{H}_{ij}(\boldsymbol{\ell}) := \braket{v_i|H_{\boldsymbol{\ell}}|v_j}$.
\item[EC overlap\\$\ecS_{ij}$] Overlap matrix between the training state vectors $\mathcal{C}_{ij} := \braket{v_i|v_j}$.
\item[Atomic Metric\\$g^{ab}$] A matrix of overlap integrals between two sets of local AO bases $\left\lbrace\phi\right\rbrace_a$ and $\left\lbrace\phi\right\rbrace_b$ for two different atomic configurations $\boldsymbol{\lambda}_a$ and $\boldsymbol{\lambda}_b$. Note that $a$ and $b$ are a label for the matrix, and not the matrix indices.
\item[EC eigenstate $\ket{\tilde{v}^{(n)}_{\boldsymbol{\ell}}}$] Approximation of the $n$-th eigenstate of the Hamiltonian at $\boldsymbol{\ell}$ within the EC representation.

\end{abbrv}

\section{Eigenvector Continuation for \emph{ab initio} calculations}
\label{sec:ECforAbInit}

The basic goal of Eigenvector Continuation (EC), also referred to as the reduced basis method (RBM) in the linear algebra community~\cite{quarteroni2015reduced,herbst2022surrogate,francis2022subspace,brehmer2023reduced}, is accessing the lowest energy solution of a family of time-independent Schr\"odinger equations which share a parametrized Hamiltonian $H_{\boldsymbol{\ell}}$
\begin{equation}
    H_{\boldsymbol{\ell}}\ket{v^{(n)}_{\boldsymbol{\ell}}} = E^{(n)}_{\boldsymbol{\ell}} \ket{v^{(n)}_{\boldsymbol{\ell}}}.
    \label{eq:ParamTISE}
\end{equation}
Given $H_{\boldsymbol{\ell}}$, the aim is to access the energies $E^{(0)}_{\boldsymbol{\ell}}$ (as well as other observables) of the ground state wave functions $\ket{v^{(0)}_{\boldsymbol{\ell}}}$ in some subset of the parameter phase space.
This is to be done without actually undertaking the exponentially expensive exact solution of Eq.~\eqref{eq:ParamTISE} for all parameter points of interest.
Instead, the aim is to approximate the ground states for an arbitrary parameter choice $\boldsymbol{\ell}$ inside the region of interest as a linear combination of a small number of selected parameter points $
\boldsymbol{\lambda}_i\in\Lambda$. 
Hence, after the exact ground state wave functions $\ket{v^{(0)}_i}$ are determined, the problem shifts to finding a set of expansion coefficients $a_i(\boldsymbol{\ell})$ such that
\begin{equation}
    \ket{v^{(0)}_{\boldsymbol{\ell}}}\approx
    \ket{\tilde{v}^{(0)}_{\boldsymbol{\ell}}} = \sum_{i\in\Lambda} a_i(\boldsymbol{\ell}) \ket{v^{(0)}_i}.
    \label{eq:ECexpansion}
\end{equation}
These coefficients can be variationally optimized by solving the corresponding generalized eigenvalue equation
\begin{align}
    \mathbf{\ecH}_{\boldsymbol{\ell}} \ket{\tilde{v}^{(0)}_{\boldsymbol{\ell}}} = \tilde{E}^{(0)}_{\boldsymbol\ell} \mathbf{\ecS} \ket{\tilde{v}^{(0)}_{\boldsymbol{\ell}}}
    \label{eq:genEig}
\end{align}
where the Hamiltonian and overlap matrix elements are
\begin{equation}
    \begin{split}
        \ecH_{ij}(\boldsymbol{\ell}) &= 
        \braket{v^{(0)}_i|H_{\boldsymbol{\ell}}|v^{(0)}_j},\\
        \ecS_{ij} &= \braket{v^{(0)}_i|v^{(0)}_j}.
    \end{split}
    \label{eq:EC_HandS}
\end{equation}
In the above equation, $\ecS_{ij}$ is in general not the identity matrix since the states $\ket{v^{(0)}_i}$ are eigenvectors of different Hamiltonians. 
\LK{In order to implement EC on a quantum computer, the Hamiltonian
and overlap matrix elements need to be measured.  This can be straightforwardly achieved via Hadamard test based circuits, as
discussed in the literature\cite{klymko2021real,cortes2022quantum,francis2022subspace}
Following that, the generalized eigenvalue problem, which is the size of the number of $\ell$ values, can be diagonalized classically.
As is the case in other subspace expansion methods based on a generalized eigenvalue problem, noise and measurement errors may lead to the condition number of the measured overlap matrix $\ecS$ growing unfavorably large.
This issue can be alleviated by performing a singular value decomposition of $\ecS$ and filtering out all singular values below some threshold~\cite{klymko2021real}.
In this work, we choose a singular value threshold of $1E-4$.}

An accurate representation of the ground states at all $\boldsymbol\ell$ of interest 
can be achieved with a judicious choice of a small number of expansion points $\boldsymbol{\lambda}_i$, which we will refer to
as training points or EC points~\cite{frame2018eigenvector,francis2022subspace}.
Such a compact representation can be of great value for phase space screening of a Hamiltonian, e.g. to characterize the existing phases and transitions.
For how to perform this ``judicious'' choice of training points, we refer to the existing literature, such as the residue estimation method presented in Ref.~\cite{herbst2022surrogate,brehmer2023reduced};
however, we note that a common ingredient in these approaches is the natural assumption that all Hamiltonians in the phase space of parameters $\boldsymbol{\lambda}$ share a single Hilbert space. This is not generally true in computational chemistry, as we discuss in detail below.

Within the context of quantum computation, the EC scheme allows a natural and potentially attractive approach to investigate the phase diagram of complex systems, where solving the Schr\"odinger Equation~\eqref{eq:ParamTISE} on the full set of parameter points to a desired accuracy is prohibitively expensive.
Indeed, if preparing the expansion states $\ket{v^{(0)}_i}$ on a quantum register is feasible, one can then measure the Hamiltonian and overlap matrix elements in Eq.~\eqref{eq:EC_HandS} and solve the generalized eigenvalue problem classically.
This strategy has been successfully demonstrated on simple spin and chemical models in Ref.~\cite{francis2022subspace}.
Of particular interest would be the application of EC to problems in \emph{ab initio} computational chemistry, where the phase space studied can be a parameterized chemical reaction.
However, a particular complication arises in the \emph{ab initio} context that needs to be addressed: the fact that the Hamiltonians for different parameter points $\boldsymbol{\lambda}_i$ will in general live, for realistic applications, in different Hilbert spaces.

\subsection{The Hilbert space problem in \emph{ab initio} computational chemistry}

When considering EC in the
context of quantum chemistry, a particular complication that
arises is the fact that the atomic basis is not necessarily
consistent for the set of training points. For example,
consider the typical task of finding a binding energy
curve of a diatomic molecule. At each separation $R$,
the atomic orbitals are centered at different points in space, which has to be
handled in computing the EC
Hamiltonian ($\ecH)$ and overlap ($\ecS$) matrices --- 
we discuss this procedure in detail in this section. A 
similar issue arises in
performing EC in finite volume 
calculations where the volume is not consistent\cite{yapa2022volume}.

\begin{figure}
    \centering
    \includegraphics[width=0.95\columnwidth]{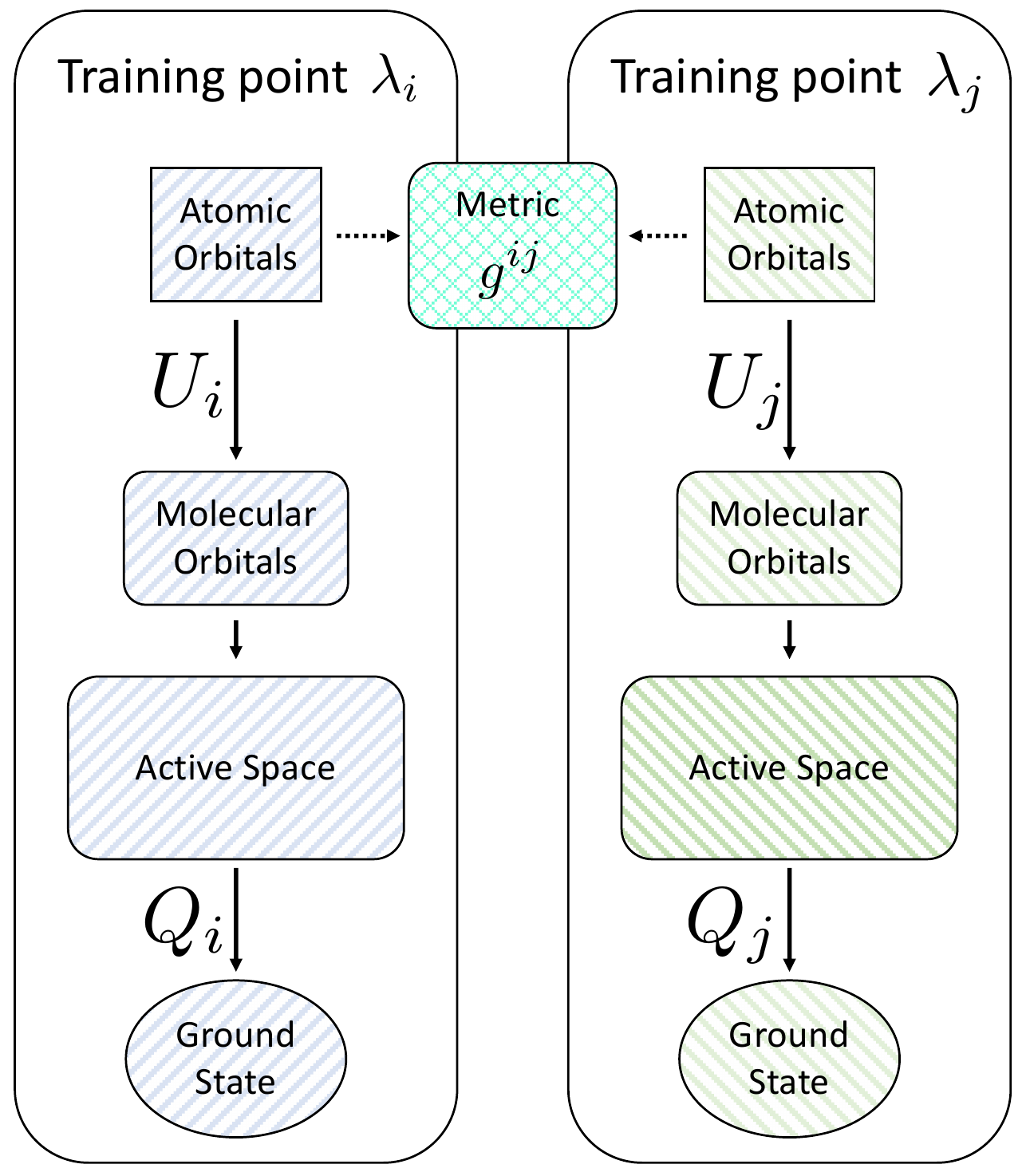}
    \caption{Diagrammatic illustration of the flow from
    atomic orbitals to the ground state in quantum chemistry. The metric $g^{ij}$ connects the atomic
    orbital bases belonging to each training point
    $\lambda_{i,j}.$ The $U$ and $Q$ matrices are rotations
    that diagonalize the Hartree-Fock and FCI
    Hamiltonians, respectively.}
    \label{fig:ec_diagram}
\end{figure}

To set the stage for our discussion, we first outline
the quantum chemistry process to obtaining a ground
state for a correlated problem (see Fig.~\ref{fig:ec_diagram}).
Each training point $\boldsymbol{\lambda}_i$ comes with a
set of atomic orbitals $\{\phi\}_i$. The initial
step of finding the ground state of the interacting
problem $\ket{v^{(0)}_i}$ is usually to
solve the Hartree-Fock problem. This is done solving the generalized
eigenvalue problem determined by the Fock-matrix ($F$)
and overlap ($\localS$) for that set of atomic orbitals (AOs).
This yields a rotation matrix $U$ which mixes the AOs
into a set of molecular orbitals (MOs).
In turn, these MOs can be used as single-particle orbital basis to define a Fock space of many-electron states in their occupation number representation.
In principle, one can then exactly solve the problem by projecting the Hamiltonian operator into this basis of Fock states, resulting in the so-called full configuration interaction (FCI) Hamiltonian $H^{FCI}$.
The FCI Hamiltonian is diagonalized via another rotation matrix $Q$ in the exponentially large Fock space to finally obtain the ground state
$\ket{v^{(0)}_i}$.
Note that in this diagonalization, one typically does not need to consider an overlap matrix, since the MOs are typically orthonormal.

One complexity arises in the EC
when the inner product must be taken between two vectors
$\ket{v^{(0)}_i}$ and $\ket{v^{(0)}_j}$ that arose from distinct
sets of atomic orbitals. This is already clear from the
overlap matrix element $\ecS_{ij}$ at the Hartree-Fock
level. Neglecting other
complexities for now, the overlap between
MOs $\ket{\alpha(\boldsymbol{\lambda}_i)}$ and $\ket{\beta(\boldsymbol{\lambda}_j)}$ that arise from the \emph{different} Hilbert spaces at training
points $\boldsymbol{\lambda}_i$ and $\boldsymbol{\lambda}_j$ is given by
\begin{align}
    \braket{\alpha(\boldsymbol{\lambda}_i)|\beta(\boldsymbol{\lambda}_j)}
    & = \sum_{mn} (U^i)^*_{n,\alpha} (U^j)_{\beta,m} \langle \phi^i_n | \phi^j_m \rangle \nonumber \\
    &= \sum_{mn} (U^i)^*_{n,\alpha} (U^j)_{\beta,m} (g^{ij})_{nm},
\end{align}
where in the last line we have introduced the
\emph{metric} $g^{ij}$ between the two training points $i$ and $j$, which is a matrix containing the inner product
between the two sets of atomic orbitals (c.f. Fig.~\ref{fig:ec_diagram}).
This already suggests that in following the EC strategy, some care will need to be taken to account for this difference in the orbital basis between different training points.
An additional step, matching the orbitals of different training points, will be necessary to evaluate the expectation values in Eq.~\eqref{eq:EC_HandS}.

To the above considerations concerning the overlap of the AOs and MOs between different geometries one has to add an additional complication which arises frequently in realistic \emph{ab initio} calculations: the notion of an active space.
Even after the massive dimensionality decrease from the uncountable real-space basis, required to describe continuous space, to the finite number of AOs, the exponential scaling of the many-body Hilbert space as a function of the number of orbitals and electrons makes it computationally impossible to perform all-orbital, all-electron calculations except in the smallest molecules with the most modest basis sets.
In all other cases, one typically restricts the post-HF (mean-field) determination of correlation effects to a subset of all orbitals, i.e. those orbitals deemed to be the most relevant for the electronic properties of the system.
These are typically chosen to be the first $N_o$ orbitals around the Fermi-level (the HOMO/LUMO frontier in the single reference description) containing the first $N_e$ electrons in the mean-field reference determinant.
These $N_o$ orbitals with $N_e$ electrons constitute the so called active space.
An effective Hamiltonian for the active space can be formulated, in which all occupied orbitals outside the active space appear only as a constant shift in energy and as modified one-body terms.
Post-HF correlated methods can then be applied to the 
active space alone, and additionally feedback correlation effects between the active space and the non-active orbitals can also be taken into account at different levels~\cite{Helgaker2000,szabo2012modern}.

\begin{figure*}
    \centering
    \includegraphics[clip=true, trim = 10 50 40 40,width=0.95\textwidth]{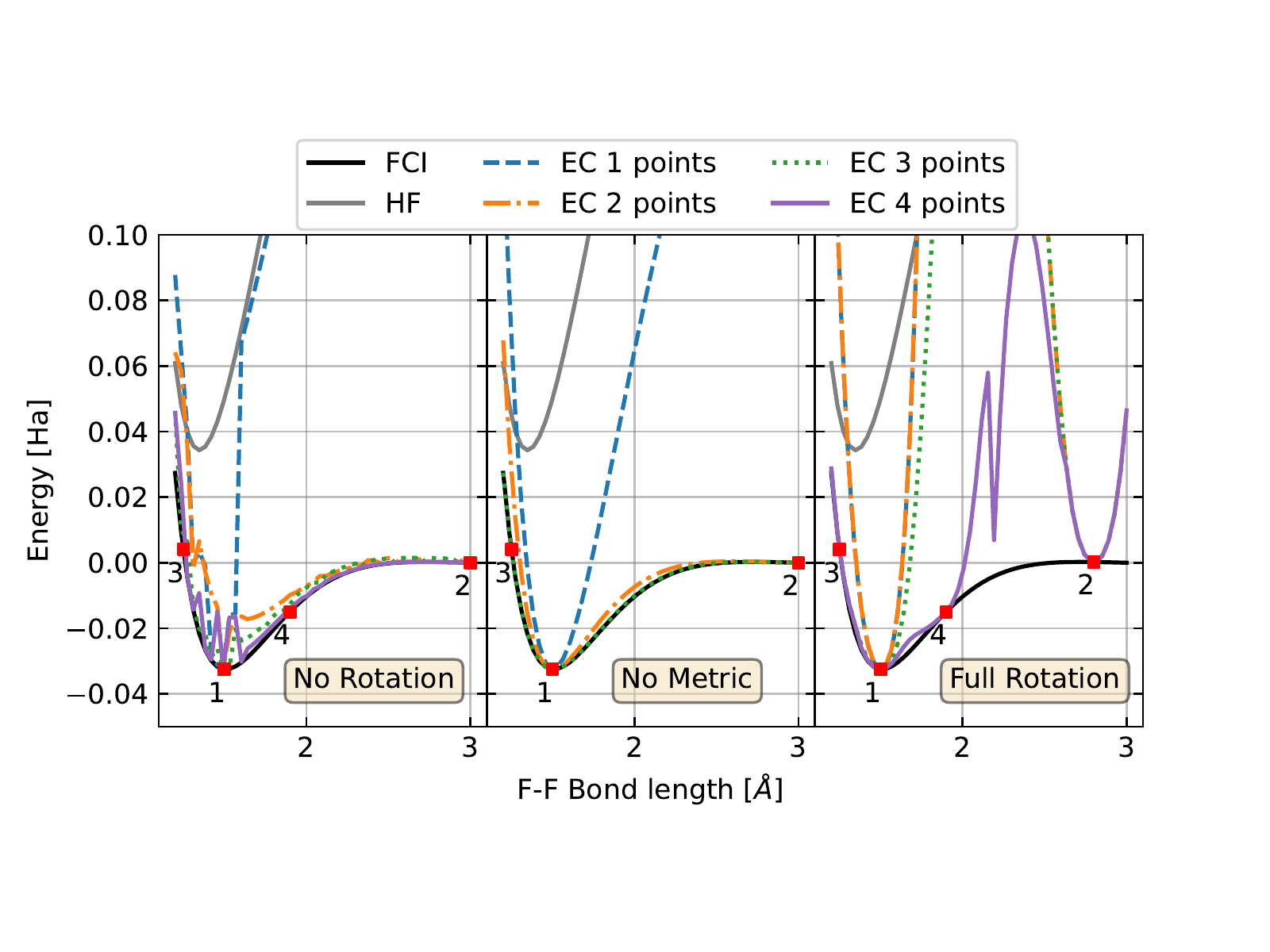}
    \caption{Potential energy surfaces (PES) for the F$_2$ dissociation using the cc-pvDZ basis and an (8o, 14e) active space. Solid lines show the Hartree-Fock (grey) and complete active space (CAS) FCI results, while the discontinuous lines present eigenvector continuation (EC) results with a different number of training points. The training points are shown as red, square markers and are labeled with numbers, representing the order in which they were included in the EC calculation. Results are shown for 3 types of orbital matchings: one ignoring all orbital rotations (leftmost), one ignoring the metric, but including all molecular orbital rotation factors (center), and finally one including all effects of the change in atomic orbital (AO) basis between geometries (rightmost).  }
    \label{fig:orb_matchings}
\end{figure*}

This notion of active space adds another layer of inconsistency between the training FCI vectors on each geometry:
Since the simplest way to define active space orbitals is in reference to the mean-field orbitals, and these change between each geometry, there is no guarantee that any given subset of them (such as the active space) span the same region of the one-body Hilbert space at each geometry.
In an extreme example, if the mean-field orbitals close to the Fermi-level have a completely different AO character between two given parameters, then the FCI vectors obtained from the corresponding effective Hamiltonians will be essentially orthogonal.

These notions of orbital matching has to be included in the evaluation of the matrix elements in Eq.~\eqref{eq:EC_HandS}, where a transformation between the now active orbital basis in $\{\ket{v^{(0)}_i},\ket{v^{(0)}_j}\}$ and $H(\boldsymbol{\ell})$ has to be performed.
If the mismatch between the active orbital basis between the EC points and the target point $\boldsymbol{\ell}$ is large, then this transformation will result in a reduction of the norm of the training vectors in the new basis.  This is detrimental to the information contained in the overlap matrix, and thus to the conditioning of the generalized eigenvalue problem, although the issue can
in principle be remedied by incorporating more training
points
to faithfully model the parameterized ground states 
in the parameter range of interest.

In order to minimize this effect, it is necessary to ensure that the active spaces in the parameter range to be studied with EC are spanned by AOs of the same character.
This can be achieved by choosing large active spaces, such that all the relevant AOs for all parameter points will always be included;
alternatively, one can choose the nature of the active space orbitals by more systematic means than proximity to the Fermi-level, e.g. using complete active space self-consistent field (CASSCF) orbital optimization (cf. Ch. 12 in Ref.~\cite{Helgaker2000}).
In the result section below, we exemplify the first strategy for weakly-correlated molecules and employ the second for the strongly correlated \ch{Cr2}.
On a molecular torsion example, we will show a case in which this orbital mismatch is harder to solve, and consequently an increased number of training points is needed to cover the full parameter space of interest.

\subsection{Possible orbital matchings}
\label{sec:orb_match}

As discussed above (Fig.~\ref{fig:ec_diagram}), finding the EC Hamiltonian and
overlap matrices involves a local rotation from
atomic orbitals (AOs) to molecular orbitals (MOs) $U$,
a local rotation from MOs to FCI eigenvectors $Q$, and an
inner product between two separate sets of atomic orbitals
(which we encode in the metric $g$).  In principle,
to capture the proper inner products, each of these
must be taken into account; in practice, however, it
can be beneficial to neglect one or more of these.

In Fig.~\ref{fig:orb_matchings} we show the results
of applying eigenvector continuation to the dissociation
of \ch{F2} using up to 4 training points and different orbital matching strategies. We compare
the binding energy to the Hartree-Fock (HF) results,
as well as a reference CAS-FCI result.
The calculations are performed in the cc-pVDZ basis set, with an (8o, 14e) active space.
The panels show three possible orbital matching approaches:
\begin{enumerate}
    \item Full rotation: The $\ecH$ and $\ecS$
    matrices are determined as discussed above, incorporating the $U$ and $Q$ rotations, as well as
    the metric.
    \item No metric: The $U$ and $Q$ rotations from the FCI vectors to the atomic orbitals are kept,
    but the metric is neglected.
    \item No rotation: The FCI vectors are treated entirely
    without reference to their origin. The $U$ and $Q$ rotations are neglected, as well as the metric.
\end{enumerate}

\begin{figure}[htpb]
    \centering
    \includegraphics[width=0.45\textwidth]{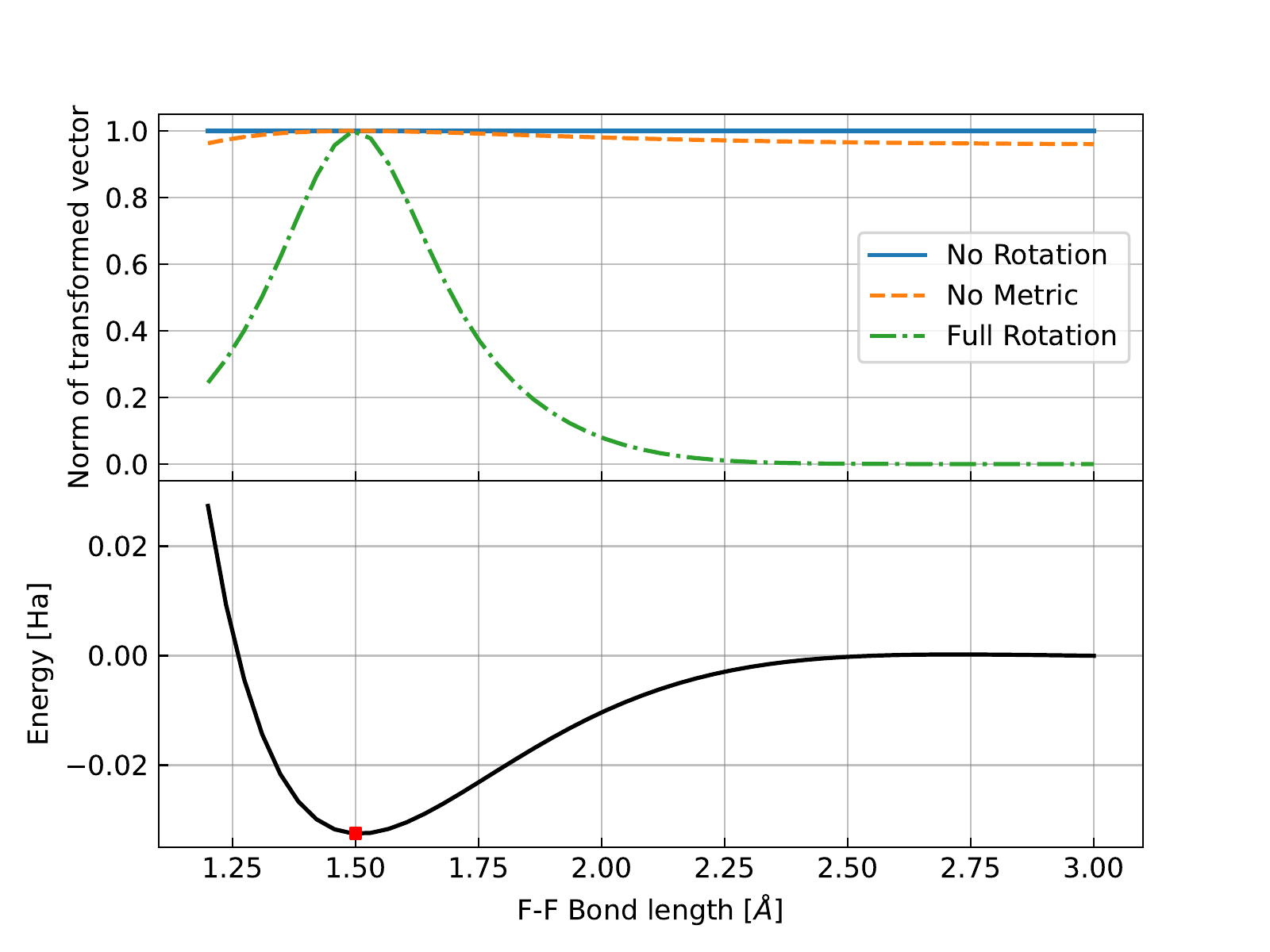}
    \caption{Norm of the transformed vector used as basis in the eigenvector continuation (EC) calculation, for the three different orbital matchings (upper panel), in an F$_2$ cc-pvDZ calculation with an (8o, 12e) active space. The original vector is the CAS-FCI solution at $R = 1.5\ \AA$ (red marker in lower panel). For reference, the FCI potential energy surface is shown as a black solid curve (lower panel). }
    \label{fig:norms}
\end{figure}

\LK{Explicit mathematical expressions for the implementation of these three orbital matching choices are presented in the Appendix.} Somewhat counter-intuitively, the best results are obtained
when the metric is neglected, and the method still works
when the metric and rotations are neglected (although
with limited success). On the other hand,
the notionally correct calculation which incorporates
the rotations and the metric performs quite poorly.

The poor behavior of the calculation with full rotations
can be understood by considering the metric. In our
calculations we use localized atomic orbitals; while
their highly localized nature is desirable from a quantum chemistry perspective, it also leads to a rapidly decaying
metric.  In essence, the overlap between atomic bases at
different training point $R$ tends exponentially to zero as $R$ increases. We illustrate this in Fig.~\ref{fig:norms} where
we plot the vector norm of one of a training state at
$R=1.5$~\AA~in the basis corresponding to a range of $R$.
When the metric is included, the norm drops and nearly
vanishes for $R > 2.5$~\AA.  The
inner product is thus not well captured, and the 
eigenvector continuation fails.

When the metric is neglected, however, eigenvector
continuation performs quite well. In particular, when
the local rotations $U$ and $Q$ are kept, 3 training
points are sufficient to obtain the full binding energy
curve. Intuitively, this can be understood as follows. The $Q$ and $U$ rotations describe how the final FCI
eigenvector is composed of the molecular orbitals,
and how the molecules orbitals are composed of the atomic 
orbitals. In other words, the FCI eigenvector at a given 
parameter $\boldsymbol{\ell}$ is a vector in the space spanned by the basis of atomic orbitals at that parameter point. 
As the atomic separation is varied the FCI eigenvector rotates in the space spanned by the local AOs. However,
it is in fact irrelevant that the local atomic
basis is now shifted in real space. For EC, it suffices that
the FCI eigenvector expressed in \emph{its own}
local basis can be spanned by the training points in
\emph{their own} local basis. This information is encoded
in $Q$ and $U$, and thus keeping those is sufficient.
Putting this together with the issues with the metric,
we conclude that neglecting the metric is a better choice
than keeping it. In Fig.~\ref{fig:norms},
we show that the previous issues with the vanishing overlap
due to the metric do not arise here.

Finally, we can choose to neglect all rotations,
and treat the FCI eigenvector as a vector divorced
from any basis information. Here, a more straightforward
linear algebra perspective is insightful. The FCI
eigenvector simply needs to be spanned by a sufficient
number of linearly independent basis vectors; the basis
vectors need to be sufficiently expressive in order to
be able to orthogonalize the ground state with respect
to any other states in the subspace. Thus, this method
works, but a larger number of training states may be required. Fig.~\ref{fig:orb_matchings} shows that
the 4 number of training points considered here
are not sufficient to achieve agreement with the
reference FCI result.

\LK{It is worth mentioning that the EC framework remains variational regardless of the orbital matching condition employed.
Indeed, the orbital matching conditions just correspond to different effective expansion bases for a given point in parameter space, but the structure of the generalized eigenvalue problem in Eq.~\eqref{eq:EC_HandS} remains the same for any of these choices.
Hence, choosing the orbital matching condition minimizing the energy, besides that which generates continuous PES, is a perfectly valid variational strategy.}

\LK{The main goal of this work concerns showing the applicability of the EC framework to \emph{ab initio} systems.
As discussed in this section, the main ingredient that need to be added to previous implementation strategies~\cite{quarteroni2015reduced,herbst2022surrogate,francis2022subspace,brehmer2023reduced} is the orbital matching between different points in parameter space.
Here, we implement this orbital matching as full orbital rotations with different rotation matrices (see Appendix) on the FCI training vectors.
This is of course not a realistic approach to target large molecular systems, since rotating the FCI vectors formally scales exponentially with system size, just like solving the FCI problem.
Future work should be dedicated to developing other, perhaps approximate orbital matching implementations circumventing the explicit FCI vector rotation. }

\section{Analyzing the Performance of Eigenvector Continuation}

\begin{figure*}
    \centering
    \includegraphics[width=\textwidth]{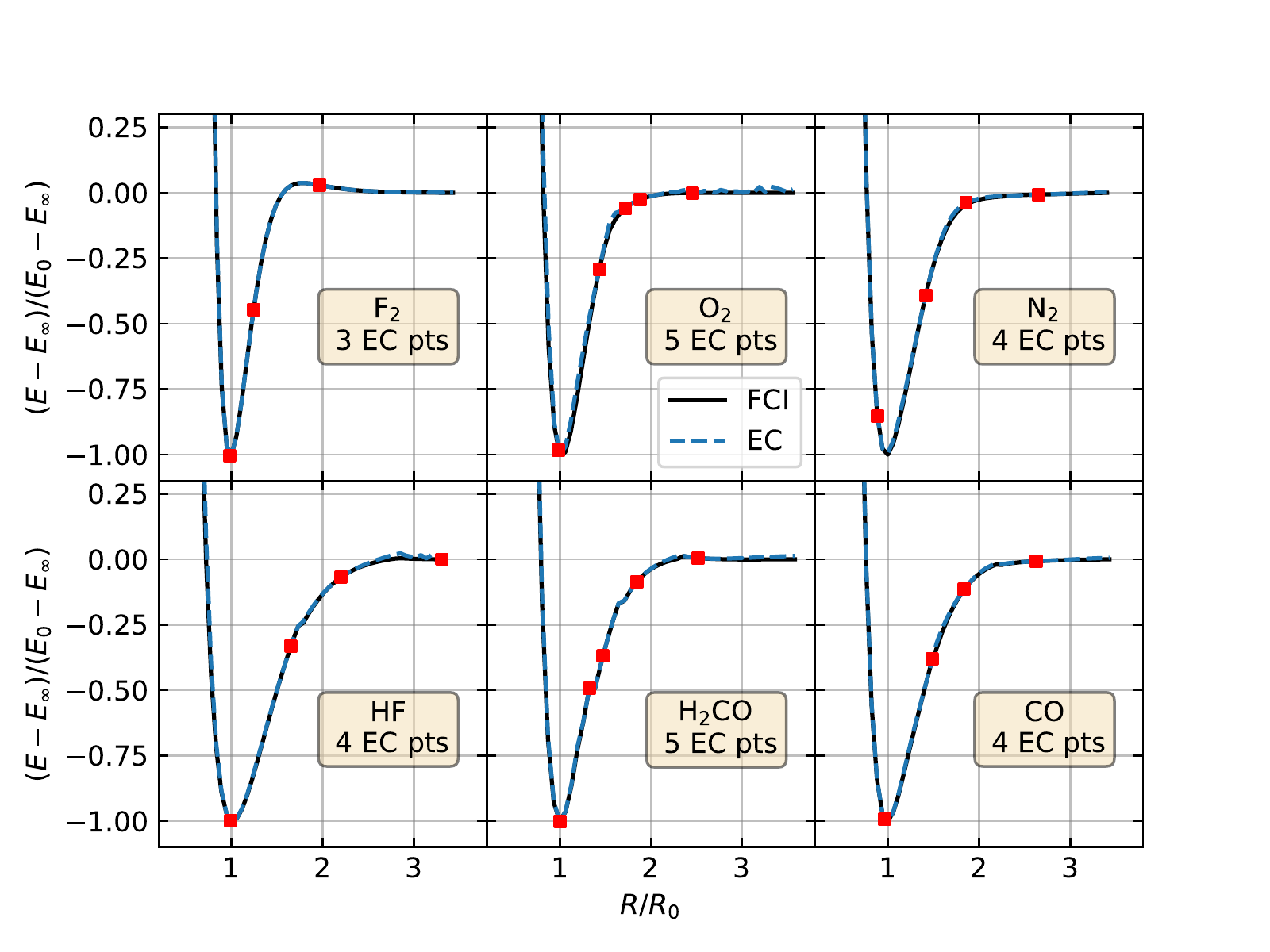}
    \caption{Bond stretching potential energy surfaces (PES) for small molecules, comparing FCI and eigenvector continuation (EC). 
    The x-axis is the bond length rescaled with the equilibrium value for the given molecule, and the y-axis is the ground state energy rescaled by the minimum value and shifted by the large distance asymptotic (i.e. the bond energy). 
    Symmetric bonds are shown in the upper row, while asymmetric bonds are found in the lower row.}
    \label{fig:ecInQc}
\end{figure*}

In this section we turn to the analysis of the reliability of EC as a compact and accurate approximation to characterize the ground state of \emph{ab initio} molecular problems in simple one-dimensional parameter spaces.
Arguably the most relevant parameter entering the molecular Hamiltonian, within the Born-Oppenheimer approximation, is the molecular geometry.
The electronic energy eigenvalues as a function of the nuclear positions are commonly referred to as potential energy surfaces (PES).
Hence, we can reformulate our goal as the study of how many EC points are necessary for accurately reconstructing one-dimensional PES in a few cases of chemical interest, namely stretching and torsion of covalent bonds.

While a one-dimensional PES for a particular molecule and bond is a fundamentally well defined target, the different approximations typically invoked in a computational chemistry calculation limit the ultimate accuracy of even a hypothetical and exact full configuration interaction (FCI) simulation.
Indeed, choices like the atomic basis set, the single-particle orbitals and the correlated active space all affect the FCI reference, and a careful analysis of the convergence of the observables of interest with respect to these factors is a necessary step in an electronic structure investigation.
However, these considerations fall beyond the scope of this work, as we concern ourselves with examining how well EC can reproduce a given FCI reference with a small number of training points.
We will thus choose a single, reasonable, but by no means final, FCI reference for each molecular case study, but make no claims as to its ultimate relevance towards the accurate description of the real physical system.
We compensate this simplification by examining molecular examples of different degree of electronic correlation and computational complexity, in order to keep the validity of our conclusions as broad as possible.

A brief description of the FCI references for all molecular systems follows, with a subsequent presentation and discussion of the numerical results.
All calculations were performed using the PYSCF package for electronic structure~\cite{pyscf1, pyscf2, pyscf3}.

\subsection{Molecular Systems}

\subsubsection{Bond Stretching of Weakly Correlated Molecules}

The majority of the PESs studied in this work fall under the category of bond stretching of ``weakly correlated'' molecules.
By this, we mean that the nature of the electronic correlation in the equilibrium geometry is well captured by single-reference methods.
Nonetheless, in the bond stretching process the ground state naturally becomes multi-reference (to some degree strongly-correlated), making the accurate description of bond dissociation energies a challenge for effective single-particle theories even in these comparatively simple molecules. In addition, the study of bond stretching
is of relevance to the quantum computing community,
where the bond stretching and dissociation problem
is a \emph{drosophila}~\cite{cao2019,lee2019,mcardle2020,googleHF,huggins2022,baek2022say,clary2023}.

We consider bonds of different chemical character.
We take into account the common heuristic distinction of single, double and triple bonds derived mostly from Lewis structures, and distinguish between symmetric and asymmetric bonds, i.e. bonds between chemically equivalent and inequivalent atoms.
As examples of single bonds, we perform EC calculations for \ch{F2} and \ch{HF}, while we consider \ch{O2} and the \ch{CO} bond in \ch{H2CO} for the double bond 
category; \ch{N2} and \ch{CO} are our triple bond representatives.
For all these systems, we used a cc-pVDZ basis set, in which at each geometry we perform a restricted Hartree-Fock (HF) calculation.
For the asymmetric bond stretchings, in order to generate smooth PES, all possible spin states within restricted open-shell HF were considered, and the lowest in energy for each bond length was used as the molecular
orbital basis for the subsequent FCI calculations.
Even in this small molecules and moderate basis set, performing FCI on all electrons and orbitals is computationally prohibitive on a single processor.
Hence, we performed instead complete active space (CAS) calculations including the 2p and 2s atomic orbital manifolds involved in the bond breaking.
We summarize the active space sizes in Tab.~\ref{tab:calcs}.
We considered bond lengths up to 3.5 times the FCI equilibrium bond length.

\begin{table}[h]
    \centering
    \begin{tabular}{|l|c|c|c|}
      \hline
      Molecule & Basis Set & Orbital Basis & Active Space \\ 
      \hline
      F$_2$   & cc-pVDZ & RHF & (8o, 14e) \\
      O$_2$   & cc-pVDZ & RHF & (8o, 12e) \\
      N$_2$   & cc-pVDZ & RHF & (8o, 10e) \\
      HF   & cc-pVDZ & ROHF & (7o, 10e) \\
      H$_2$CO   & cc-pVDZ & ROHF & (8o, 10e) \\
      CO  & cc-pVDZ & ROHF & (8o, 10e) \\
      Cr$_2$   & cc-pVTZ-dk & CASSCF & (12o, 12e) \\
      C$_6$H$_8$   & cc-pVDZ & RHF & (6o, 6e)\\
      \hline
    \end{tabular}
    \caption{Table summarizing the computational details of the molecular potential energy surfaces (PES) studied in this work.}
    \label{tab:calcs}
\end{table}

For all the 6 molecules presented in Fig.~\ref{fig:ecInQc}, the PES curve
obtained by EC is in good agreement with the FCI reference.
To ease the comparison between different molecules, we rescale the bond length axis by the FCI equilibrium distance of each molecule, and the energy axis by the bonding energy, taking the energy at 3.5 times the equilibrium bond length as the dissociated asymptote.
For each molecule, we present the minimal number of EC training 
points that leads to an acceptable result compared to FCI.
None of the molecules require a particularly large number,
but some variation does exist between the molecules. We note
in particular 
that \ch{F2}, \ch{N2} and \ch{CO} exhibit better agreement;
while the other EC PES curves have some departure from the
FCI result, these could be readily improved by the addition
of more EC points (c.f. Fig.~\ref{fig:orb_matchings}).
In the case of \ch{F2}, it is remarkable that just 3 points are enough to recover the full PES faithfully.
These can be interpreted as the three distinct physical regions in the bond dissociation process: the bound region, the dissociated region, and the Coulson-Fisher point where a mean-field description would start breaking translational and/or spin symmetry.
While for all other bonds shown in Fig.~\ref{fig:ecInQc} more than 3 training points are needed, these typically agglomerate around the Coulson-Fisher region, where the system is arguably more strongly correlated.
In this sense, a qualitative relationship can be established between the variability of the eigenstate character in a bond-length region and the number of EC points needed to sample that zone accurately.
This matches well the observations using EC in spin models~\cite{herbst2022surrogate,francis2022subspace,brehmer2023reduced}.
We note that there are some unusual kinks in the PES curves for the asymmetric bond breakings, which is due to the limitations of the FCI
calculations, rather than an artifact introduced by EC.

\subsubsection{Cr$_2$ and Bond Torsion}

\begin{figure}
    \centering
    \includegraphics[width=0.5\textwidth]{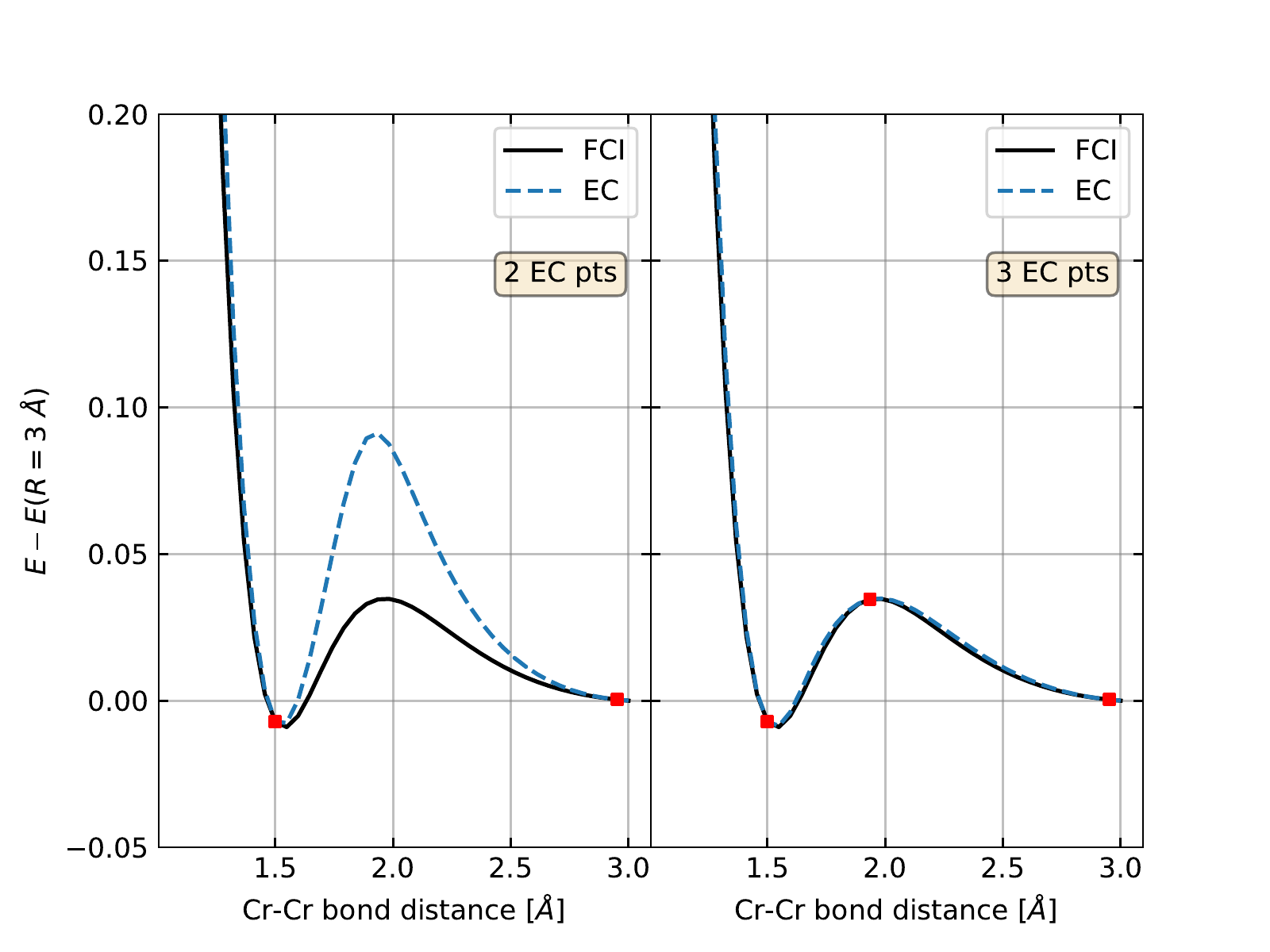}
    \caption{Potential energy surface (PES) for Cr$_2$ dimer in cc-pvTZ-dk basis. The FCI results correspond to a CASSCF (12o, 12e) calculation. Shown are eigenvector continuation (EC) for two different numbers of training points.}
    \label{fig:corr}
\end{figure}

\begin{figure*}
    \centering
    \includegraphics[width=\textwidth]{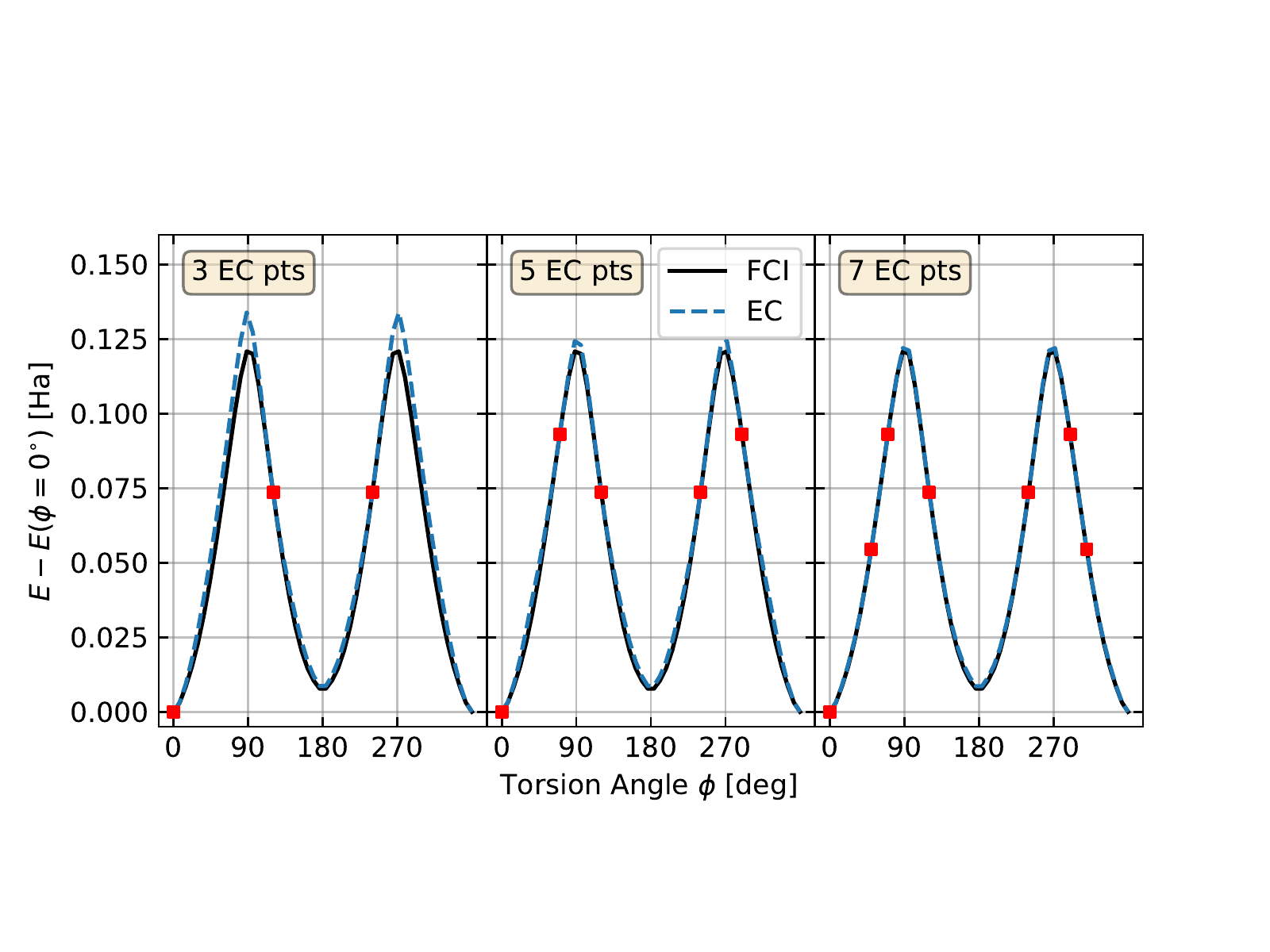}
    \caption{Potential energy surface (PES) for hexatriene in the cc-pvDZ basis as a function of the torsion angle $\phi$ around the central C-C double bond. $\phi=0^{\circ}$ corresponds to the trans configuration, $\phi=180^{\circ}$ to cis. The FCI results correspond to a complete active space (6o, 6e) calculation involving only the $\pi$ orbital manifold. Shown are eigenvector continuation (EC) for three different numbers of training points, always symmetrically chosen around $\phi = 180^{\circ}$.}
    \label{fig:torsion}
\end{figure*}

To test the performance of EC for intrinsically strongly correlated molecules, we consider the bond stretching of a Cr$_2$ dimer, where we used 
a cc-pVTZ-dk basis set.
Besides the restricted HF calculation at each bond length, a further orbital optimization was performed at the complete active space self-consistent field (CASSCF) level of theory, with a (12o, 12e) active space.
The multi-reference orbital optimization was necessary to obtain a homogeneous 3$d$ and 4$s$ orbital character in the active space through the bond dissociation.
The (12o, 12e) CASSCF energy served then also as FCI reference for the EC.  Fig.~\ref{fig:corr} shows the 
results of the EC calculation for 2 and 3 training points.
As was observed in the weakly correlated molecules, in \ch{Cr2} as well a sparse sampling of the bound, dissociated and Coulson-Fisher regions is sufficient to recover the full PES.

Considering all the bond stretching examples, it is remarkable that the EC scheme seems to be fairly insensitive to the chemical nature of the PES modelled.
Indeed, regardless of the chemical complexity of the bond, represented by single, double, triple, symmetric, asymmetric and correlated bonds, as well as the computational complexity of the FCI reference, based on either RHF or CASSCF orbitals, the EC representation shows a relatively homogeneous convergence in terms of the training points.
A handful (up to five) points along the bond stretching, typically including at least the bound, dissociated and Coulson-Fisher region, are enough to obtain a visually accurate representation of the ground state PES along the full reaction path.

Finally, we considered the bond torsion of trans-hexatriene around the central CC double bond.
This PES was evaluated in the cc-pVDZ basis, using a minimal active space including all $\pi$ orbitals, namely (6o, 6e), on restricted HF orbitals.
The $\phi = 0^{\circ}$ geometry was taken from Ref.~\cite{huggins2020non}.
The orbital mismatch problem is more severe in this case as the rotation mixes the $p$-orbital manifold. 
By rotating around the bond, the atomic $p_z$-orbitals of one half of the molecule become eventually completely orthogonal to the $p_z$ orbitals of the other half, and consequently the AO character of the frontier MOs changes drastically from 0 torsion to 90$^\circ$.
As a result, the PES
requires a larger number of training points (7) to capture the full surface properly.
Nonetheless, this is still  modest sampling with which to recover the full PES.

\begin{figure*}[t]
    \centering
    \includegraphics[width=\textwidth]{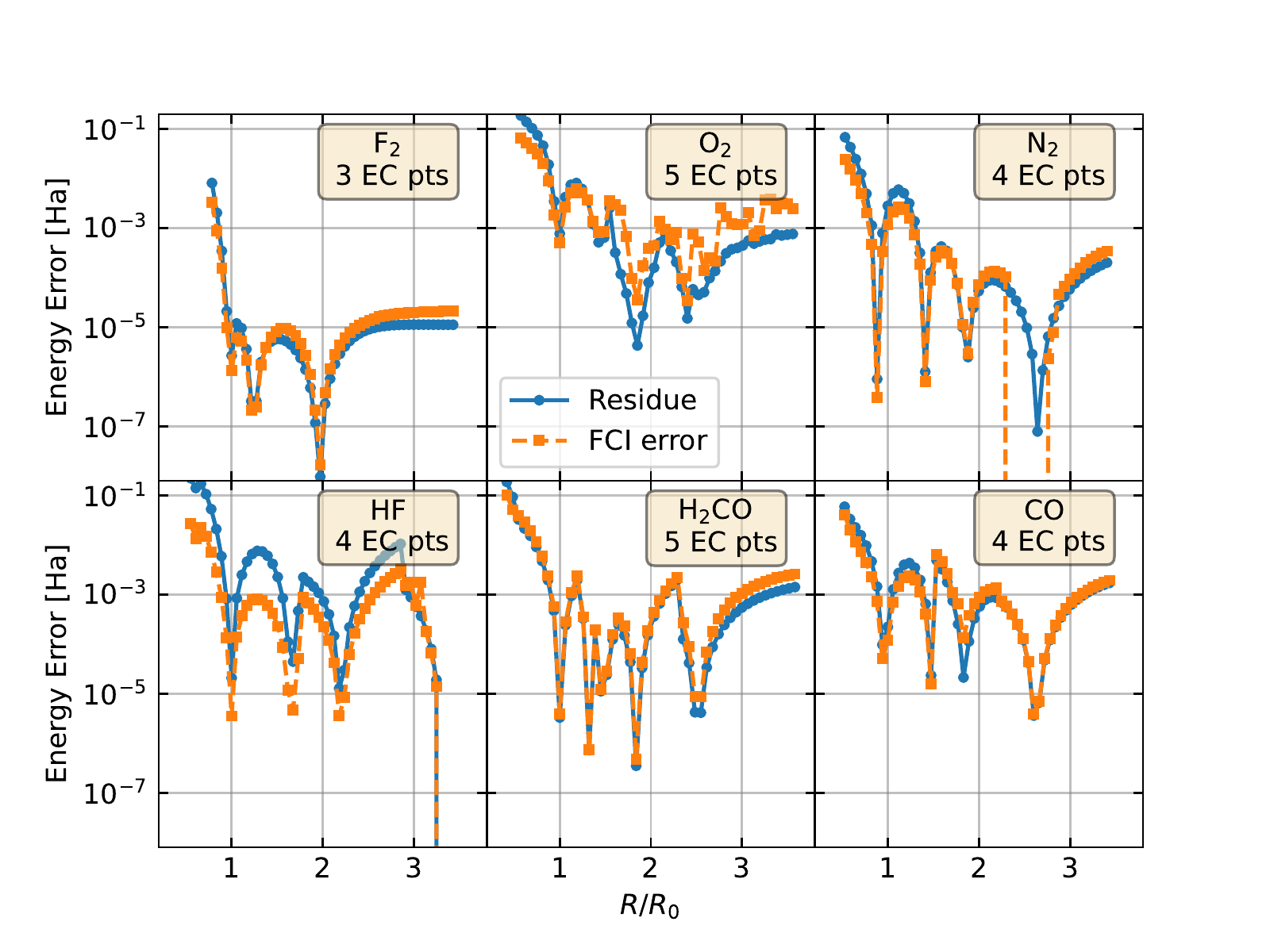}
    \caption{Two measures of the error on the potential energy surfaces (PES) for the bond stretching examples in Fig~\ref{fig:ecInQc}. The exact error with respect to FCI is shown with square markers, and the residue estimate in Eq~\eqref{eq:residue} with round markers. The approximate residue follows the exact error closely.}
    \label{fig:residues}
\end{figure*}

\subsection{Choosing the EC training points}

As mentioned in Sec.~\ref{sec:ECforAbInit}, how to judiciously choose the EC training points to maximize the compactness of the approximation without compromising accuracy has been discussed by Herbst et al.~\cite{herbst2022surrogate},
who suggest the use of a residual estimate for determining what points to add. In essence, given a previous EC approximation, the next point to
be chosen is the one that maximizes the additional information in the basis as measured by the accuracy
of the EC eigenvalue at that point.
Here, we briefly exemplify how the residue estimate proposed therein satisfactorily adapts to the aim of achieving ``chemical accuracy'' in \emph{ab initio} simulations.
By chemical accuracy, one refers to a maximal error of 1.6~mHa of a computational estimate with the true or reference value.
When controlled experimental results are not available, one often uses as reference a computational result from a more accurate theoretical model.
For our purposes, we can use the FCI reference to determine the error of our EC results at each molecular geometry.

If the average error in the energy of a given PES curve calculated via EC with $m$ training points ($m$-EC) is above chemical accuracy, a natural choice for the $m+1$-th training point is to pick a molecular geometry in the region of maximal deviation.
The FCI reference is in general not available, and it is necessary to obtain an estimate of the error using exclusively the data available within the EC calculation.
Following Ref.~\cite{herbst2022surrogate}, one can evaluate the residue of the EC approximation at each geometry of interest.
Given a geometry $\boldsymbol{\ell}$, for which the $m$-EC simulation provides with a ground state wave function approximant $\ket{v^{(0),m}_{\boldsymbol{\ell}}}$ with estimate energy $\tilde{E}^{(0),m}_{\boldsymbol{\ell}}$, the residue is defined as

\begin{equation}
    r^{(m)}_{\boldsymbol{\ell}} = \left|H_{\boldsymbol{\ell}}\ket{v^{(0),m}_{\boldsymbol{\ell}}}-\tilde{E}^{(0),m}_{\boldsymbol{\ell}}\ket{v^{(0),m}_{\boldsymbol{\ell}}}\right|^2.
    \label{eq:residue}
\end{equation}

This residue $r^{(n)}_{\boldsymbol{\ell}}$ can be written in terms of the Hamiltonian and overlap matrices, the eigenvector from the generalized eigenvalue problem in Eq.~\eqref{eq:genEig}, and the matrix elements of the squared Hamiltonian in the EC training basis $(H^2)_{ij} = \braket{v^{(0)}_i|H^2_{\boldsymbol{\ell}}|v^{(0)}_j}$.
The only additional cost on top of the EC calculation is thus the measurement of the squared Hamiltonian matrix elements.

In Fig.~\ref{fig:residues}, we present the residues of the EC calculations for the bond stretchings in Fig.~\ref{fig:ecInQc}, keeping the same number of training points.
We compare these to the error in the energy with respect to the FCI energies in the corresponding active spaces.
As can be seen in Fig.~\ref{fig:residues}, the residue estimate closely follows the actual error with respect to FCI, and thus offers an effective indicator to choose the EC training points in order to ultimately reach chemical accuracy with respect to the FCI reference.
Of course, this does not guarantee an excellent agreement with experiment, as several approximations enter the FCI reference chosen in each case.
However, the compactification offered by the EC approximation enables FCI-quality results within a small fraction of the cost of actually performing an FCI-level calculation (be it using a classical or quantum algorithm) at each point on the potential energy surface.

\begin{figure*}[htpb]
    \centering
    \includegraphics[width=\textwidth]{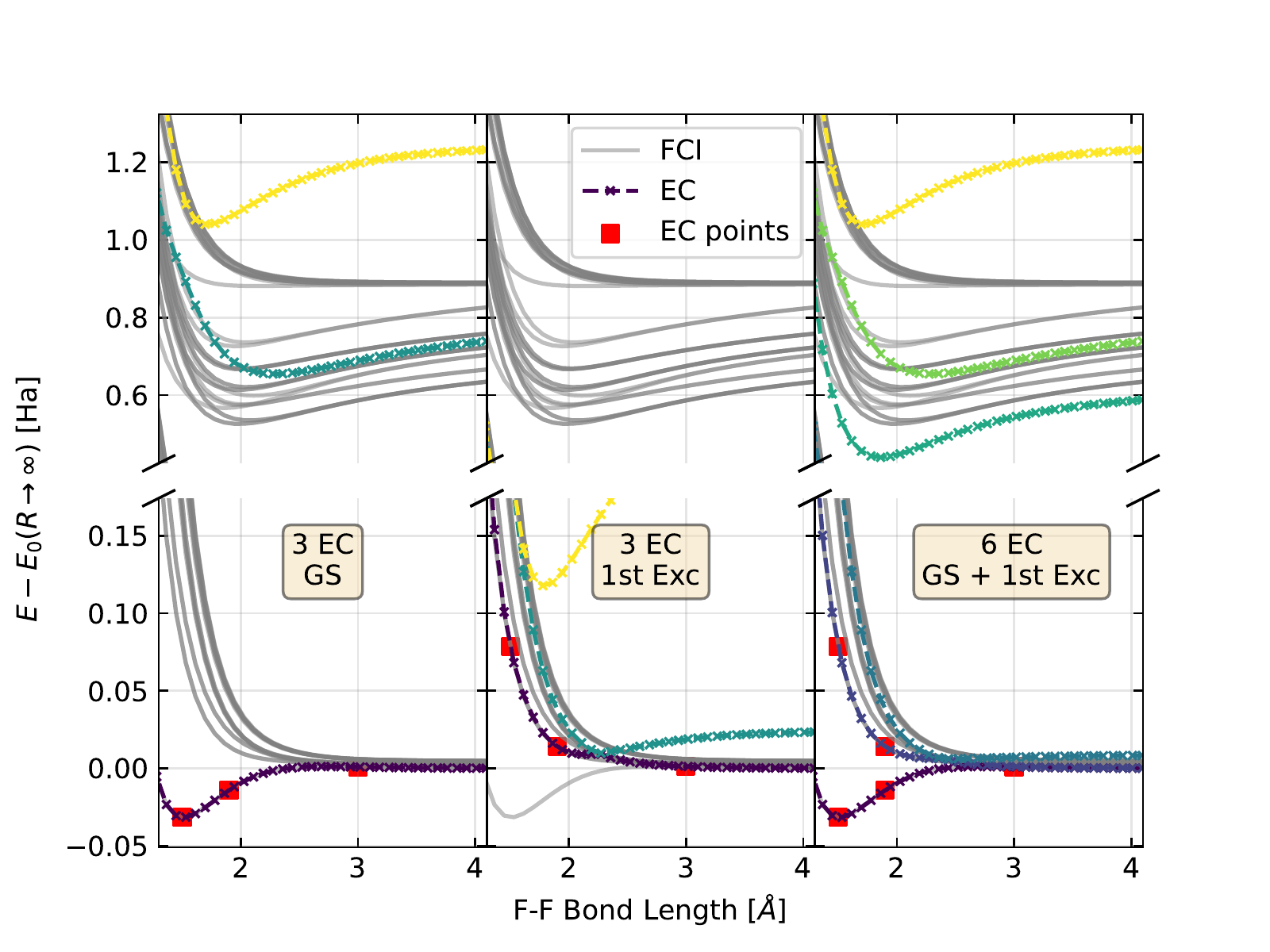}
    \caption{Results of eigenvector continuation (EC) for excited state potential energy surfaces (PES) in F$_2$ using the cc-pVDZ basis set. The FCI PES in the (8o, 14e) active space for the first few excited states are presented in grey. Results are shown for three different EC simulations. Left panel: EC with 3 training points using always FCI ground state vectors. Middle panel: EC with 3 training points using always FCI first excited state vectors. Right panel: EC with 6 training points in 3 different geometries, using both the ground state and 1st excited state of the FCI Hamiltonian in each point.}
    \label{fig:F2_exc}
\end{figure*}

\subsection{Accessing excited states with EC}

In principle, the EC formalism is not limited to the approximation of ground states.
As long as more than one training point is used, the generalized eigenvalue problem in Eq.~\eqref{eq:genEig} will have multiple eigenvectors, some of which could be accurate approximations of some excited state in the full Hamiltonian.
Indeed, there are two scenarios in which it is natural to expect EC to provide a compact representation of excited states.
First, consider Hamiltonians that have avoided level crossings, such that the ground and first excited states switch character continuously across some path in the parameter space.
In this case, using only ground states as training points along a path through the level crossing should also potentially result in an acceptable representation of the first excited state.
Second, there is no fundamental need to use exclusively FCI ground states to build the EC training sets.
If excited states are used, this should produce an EC approximation targeting the corresponding excited state PES.
Simultaneously, in the presence of the aforementioned level crossings, having a mixed EC training set containing ground and excited states can lead to accurate representations for both.
Here we consider these different possibilities in the example of the F$_2$ dimer.

We present excited state PES for the F$_2$ molecule in the cc-pVDZ basis, with a (8e, 14o) active space in Fig.~\ref{fig:F2_exc}.
The FCI surfaces, shown as grey lines, do not show a level crossing between the ground and first excited state along the dimer bond dissociation.
Nonetheless, these states become degenerate in the dissociation limit, and hence a complete decoupling between both states is not obvious \emph{a priori}.
Fig.~\ref{fig:F2_exc} shows the results for three different EC calculations in three panels.
For all of these, the training points were obtained from the same molecular geometries, matching those already shown in the upper left panel of Fig.~\ref{fig:ecInQc}.

In the left-most panel of Fig.~\ref{fig:F2_exc}, we compare all three eigenstates obtained from the effective Hamiltonians of a ground state based EC to the first few lying FCI eigenstates.
While the exact ground state PES is perfectly reproduced, as was already discussed in Fig.~\ref{fig:ecInQc}, the excited states of the effective EC Hamiltonian do not match well any of the exact excited state PESs.
Moreover, these EC excited state PESs come after the first group of close-lying FCI excited state PESs, appearing at $\sim0 .8$~Ha above the ground state.
This suggests that the subspace that captures the ground state PES could be orthogonal to the first bundle of excited states.

A similar picture occurs in the middle panel of Fig.~\ref{fig:F2_exc}, where the three training points in the EC simulation were all first excited states.
Consequently, a relatively faithful approximation of the FCI first excited state PES is obtained, although a noticeable deviation appears at $\sim 2.2~\AA$ through an artificial, i.e. not present in the FCI reference, avoided crossing with the second excited state.
Similarly to the first case, neither of the two higher lying excited states from the EC calculation approximate any of the FCI PESs well.
Surprisingly, all the EC curves in this case have some degree of bonding character (a minimum), even when all excited states in the corresponding energy window are dissociating, including the one used to obtain the training points. 
Still, since there is no overlap with the results from the EC calculation using ground state training vectors, it seems that the ground state and first excited eigenstate manifolds are mostly decoupled.

To further confirm this, an EC calculation was performed using 6 training points in the same 3 molecular geometries, shown in the right-most panel of Fig.~\ref{fig:F2_exc}.
For each of the 3 bond lengths, the ground and first excited states at the FCI level were used as independent training vectors for the EC simulation.
The three PES already obtained in the EC calculation using only ground state training points (cf. left panel in Fig.~\ref{fig:F2_exc}) are found again in this larger simulation.
However, the three PES that would correspond to the EC calculation based on excited states only (cf. middle panel in Fig.~\ref{fig:F2_exc}) are significantly changed.
The lowest in energy of the three --- the overall first excited state of the EC simulation --- follows the exact result better than in the middle panel, missing the deviation at $\sim2.2$\AA.
Furthermore, the next excited state PES is significantly shallower, closer to the expected non-binding behavior.
Finally, the highest excited state among the central three is now pushed in energy much closer to the excited state manifold at $\sim 0.8$~Ha, better justifying its bound character.
Despite these noticeable improvements, we still find that the only PES that accurately reproduces the FCI reference are those which are sampled explicitly, namely the FCI ground and first excited state in this case.

\section{Conclusions}

The spirit of the eigenvector continuation (EC) approach is proposing low-dimensional effective models to accurately reproduce targeted eigenstates of a parameterized Hamiltonian in some region of the parametric phase space.
This is done by sampling a small number of points in said region, i.e. performing a computationally expensive, accurate determination of the eigenstates of interest in these few points, and then using their information to reconstruct the eigenstates inexpensively in the rest of phase space.
The computation at the training points may be exact full configuration interaction (FCI) when feasible~\cite{herbst2022surrogate}, based on highly accurate matrix-product state (MPS) Ans\"atze~\cite{brehmer2023reduced}, or even the result of a quantum computation~\cite{francis2022subspace} for systems beyond the current reach of classical approaches.
With a modest number of training points, the accurate results of comparable quality to these expensive methods can be recovered in the full parameter phase space at a fraction of the computational complexity.
This becomes especially attractive for studying chemical reactions, which involves the accurate determination of the potential energy surfaces (PES) of ground and excited states along the reaction coordinates.
Therefore, here we have investigated the applicability and effectiveness of EC in the \emph{ab initio} setting.

One of the major hurdles in applying EC to \emph{ab initio} quantum chemistry is the
mismatch in basis that arises from disparate molecular geometries for the subspace basis
point, which we discussed extensively in the text.  One significant conclusion from this
work is that parts of this mismatch may be entirely neglected; specifically, the mismatch
between the most basic level of the calculations, i.e. the atomic orbital overlap
between different training states.  After doing so, we have shown that the PES can be
captured with remarkably few subspace basis vectors for a
number of chemically distinct molecules single, double and triple bonds between chemically equivalent and inequivalent atoms in weakly correlated molecules, bond stretching of the intrinsically strongly correlated \ch{Cr2}, and the bond torsion of trans-hexatriene around the central CC double bond.  The associated error as compared to the FCI reference calculations
is quite low.

Several aspects of the results that go beyond simple ground state
manifolds are worth highlighting.  First, EC can correctly
handle level crossings in the ground state spectrum in chemical
molecules, as long as training points are chosen on both sides of the
crossing;  this extends to any situation where multiple orthogonal subsectors
are of interest.  Second, 
we have shown that the use of eigenvector continuation is not limited to the
ground state.  Excited state manifolds can also be captured by inclusion of representatives
of the excited states into the subspace. We exemplify this in \ch{F2} by sampling with excited states instead of ground states.

\LK{There are two promising directions of future work on the EC framework worth mentioning at this point: 
First, as discussed in Sec.~\ref{sec:orb_match}, the current implementation of EC, which involves rotating the exponentially large FCI vectors, is not suitable for large calculations. 
Thus, in order to extend its impact to complex PES in larger molecules, there is need to develop an alternative approach to evaluate the expectation values in Eq.~\eqref{eq:EC_HandS}, heeding the issues with orbital matching presented here but avoiding the explicit rotation of the FCI vectors. 
The other direction concerns the use of approximate solutions instead of exact FCI for the training points.
Indeed, any approximate ansatz giving access to the expectation values in Eq.~\eqref{eq:EC_HandS}, upon performing an orbital matching, can be used to perform the EC scheme. Moreover, the use of such approximate states does not affect the variational property of the resulting PES, just the obtained accuracy. 
When using EC as a classical algorithm, one could consider employing coupled cluster based approximations~\cite{Bartlett2007}, while in a quantum algorithm adiabatic state preparation~\cite{Veis2014,Kremenetski2021} or other subspace expansion algorithms~\cite{parrish2019quantum,stair2020multireference,klymko2021real} could be employed.
In either case, it is interesting to note that the PES obtained using EC at the training points themselves can be variationally more accurate than the approximate solution it is built from.
In this sense, EC can be a way of not only extracting the most information out of a small number of accurate PES samples, but also of improving the accuracy of said samples.}

In short, eigenvector continuation is a promising tool for
\emph{ab initio} calculations in any situation where the eigenstates are difficult to obtain.  This is in particular true on quantum
computers, where finding ground states is a primary target and yet
remains elusive; the current state of the art is plagued with issues
in the optimization.  It is thus quite difficult to find a ground state,
and when this feat is accomplished, it should be used to maximum effect.
Eigenvector continuation is one way to achieve this goal.

\section*{Acknowledgements}
CMZ acknowledges financial support from the European Research Council (ERC), under the European Union’s Horizon 2020 research and innovation programme, Grant agreement No. 692670 ”FIRSTORM”.
AFK acknowledges financial support from the US National Science Foundation
under grant no. NSF DMR-1752713

\bibliographystyle{apsrev4-2}

\clearpage
\onecolumngrid
\appendix

\section{Equilibrium geometries for \ch{H2CO} and trans-hexatriene}

The equilibrium geometry used for \ch{H2CO} in this paper was optimized using the PYSCF interface to PyBerny~\cite{hermann2016} at the restricted Hartree-Fock level, using the cc-pVDZ basis.
The obtained geometry is presented in Tab~\ref{tab:h2co_geom}.
The bond stretched in Fig.~\ref{fig:ecInQc} is the CO double bond.

\begin{table}[h]
    \centering
    \begin{tabular}{|l|r|r|r|}
         \hline 
         Atom &  X & Y & Z \\
         \hline 
         C & 0.000000 & 0.000000 & 0.000000 \\
         O & 0.000000 & 0.000000 & 1.181970 \\
         H & 0.000000 & -0.932542 & -0.586845 \\
         H & 0.000000 & 0.932542 & -0.586845 \\
         \hline
    \end{tabular}
    \caption{Equilibrium geometry for\ch{H2CO} as used for the bond stretching in Fig.~\ref{fig:ecInQc} and Fig.~\ref{fig:residues}.}
    \label{tab:h2co_geom}
\end{table}

The equilibrium geometry of trans-hexatriene (\ch{C6H8}), i.e. the geometry corresponding to $\phi = 0^{\circ}$ in Fig.~\ref{fig:torsion}, was taken from Ref.~\cite{huggins2020non}. We reproduce it in Tab.~\ref{tab:c6h8_geom} for completeness. The rotation in the paper is performed around the CC-bond between the carbon atoms in the first and third lines of Tab.~\ref{tab:c6h8_geom}.

\begin{table}[h]
    \centering
    \begin{tabular}{|l|r|r|r|}
         \hline 
         Atom &  X & Y & Z \\
         \hline 
         C & 0.5987833 & 0.2969975 & 0.0000000 \\
         H & 0.6520887 & 1.3822812 & 0.0000000 \\
         C & -0.5987843 & -0.2970141 & 0.0000000 \\
         H & -0.6520904 & -1.3822967 & 0.0000000 \\
         C & -1.8607210 & 0.4195548 & 0.0000000 \\
         H & -1.8010551 & 1.5036080 & 0.0000000 \\
         C & -3.0531867 & -0.1693136 & 0.0000000 \\
         H & -3.9685470 & 0.4053361 & 0.0000000 \\
         H & -3.1479810 & -1.2485605 & 0.0000000 \\
         C & 1.8607264 & -0.4195599 & 0.0000000 \\
         H & 1.8010777 & -1.5036141 & 0.0000000 \\
         C & 3.0531816 & 0.1693296 & 0.0000000 \\
         H & 3.9685551 & -0.4052992 & 0.0000000 \\
         H & 3.1479561 & 1.2485793 & 0.0000000 \\
         \hline
    \end{tabular}
    \caption{Equilibrium (i.e. $\phi = 0^{\circ}$) geometry for trans-hexatriene, as reported in Ref.~\cite{huggins2020non}.}
    \label{tab:c6h8_geom}
\end{table}

\section{Derivation of the generalized eigenvalue equation}

\LK{
In this section, we briefly review the mathematics that leads to the 
generalized eigenvalue problem in Eq.~\ref{eq:genEig}.  We start
with the usual formulation of an eigenvalue problem --- for a linear operator $\ham$, an eigenvector $\ket{v}$ satisfies
\begin{align}
    \ham \ket{v} = \lambda \ket{v}.
\end{align}
We can turn this into a matrix equation by expanding $\ket{v}$ in an orthonormal basis:
\begin{align}
    \ket{v} = \sum_j v_j \ket{j}.
\end{align}
Then,
\begin{align}
    \bra{i} \ham \ket{v} &= \lambda \braket{i|v} \\
    \sum_j \braket{i|\ham|j} v_j &= \lambda \sum_j v_j \braket{i|j} \\
    \sum_j \ham_{ij} v_j &= \lambda v_i
\end{align}
which is the usual matrix eigenvalue equation and we have used the fact that $\braket{i|j} = \delta_{i,j}$.  A generalized eigenvalue
problem arises when the basis $\left\{ \ket{i} \right\}$ is not orthogonal. Then $\braket{i|j}$ is the $(i,j)$-th element of an
overlap matrix $\mathcal{C}$, and we obtain
\begin{align}
    \sum_j \ham_{ij} v_j &= \lambda \sum_j \mathcal{C}_{ij} v_j
\end{align}
}

\section{Mathematical expressions for orbital matching conditions}

\LK{In this section, we briefly motivate the different orbital matchings presented in Sec. IIIB and Fig.~\ref{fig:orb_matchings}, and give explicit mathematical expressions for the transformation matrices.
For the sake of completeness, we start reviewing orbital bases. 
Given a basis of $N_{AO}$ non-orthogonal atomic orbitals (AO) at the parameter space point $\boldsymbol{\lambda}_i$ (eg. a given molecular geometry), denoted by $\left\{\ket{\phi^i_m}\right\}_{m=1}^{N_{AO}}$, their local overlap integrals are collected in the matrix $S_i$ as
\begin{equation}
    S_{i;mn} = \braket{\phi^i_m|\phi^i_n}.
    \label{eq:SI_ovlp}
\end{equation}
Usually, the first step in a quantum chemistry computation involves finding an optimal molecular orbital (MO) basis (eg. in the mean-field sense for Hartree-Fock), which can be expressed as a linear combination of atomic orbitals, such as
\begin{equation}
    \ket{\alpha(\boldsymbol{\lambda}_i)} = \sum_m U^i_{\alpha m} \ket{\phi^i_m}.
    \label{eq:SI_MOs}
\end{equation}
These MOs are typically chosen orthonormal, ie. $\braket{\alpha(\boldsymbol{\lambda_i})|\beta(\boldsymbol{\lambda}_i)}=\delta_{\alpha\beta}$, such that the matrices $U^i$ and $S_i$ fulfill
\begin{equation}
    U^i S_i\left(U^i\right)^\dagger = \mathbb{I}.
    \label{eq:SI_orthoMO}
\end{equation}
Hence, the inverse of $U^i$ is the matrix $\sqrt{S_i}\left(U^i\right)^\dagger$ is unitary.
Given two different MO bases at one parameter point $\boldsymbol{\lambda}_i$, with corresponding ``AO to MO'' matrices $U^i$ and $W^i$ respectively, then the orbital transformation between these two MOs $T^i_{U\rightarrow W}$ is given by
\begin{equation}
    T^i_{U\rightarrow W} = W^i S_i \left(U^i\right)^\dagger = W^i \sqrt{S}_i \sqrt{S}_i \left(U^i\right)^\dagger,
    \label{eq:SI_localMOtrafo}
\end{equation}
which is unitary since both MO bases are orthonormal.
In order to transform a many-electron object between MO bases, such as the full molecular wave function, one has to apply the exponential operator $\exp\left[(\vec{c})^\dagger T^i_{U\rightarrow W}\vec{c}\right]$, where $(\vec{c})^\dagger$ is a vectorized representation of the list of creation operators in the MO basis of $U^i$.
}

\LK{Now, we consider two MOs corresponding to \emph{different} parameter points $\boldsymbol{\lambda}_i$ and $\boldsymbol{\lambda}_j$, with their ``AO to MO'' matrices $U^i$ and $W^j$ respectively.
Each parameter point has further its own AOs with their own local overlap matrices $S_i$ and $S_j$.
The non-local overlap between the AOs at both parameter points also become relevant in this scenario.
These form the metric $g^{ij}$ introduced in the main text, namely
\begin{equation}
    \left(g^{ij}\right)_{nm} = \braket{\phi^i_n|\phi^j_m}.
    \label{eq:SI_metric}
\end{equation}
This metric enters the overlap between MOs from the two different parameter points, such as
\begin{equation}
    \braket{\alpha(\boldsymbol{\lambda}_i)|\beta(\boldsymbol{\lambda}_j)} = \sum_{mn}\left(U^i_{\alpha m}\right)^*\left(g^{ij}\right)_{mn}W^j_{\beta n}.
    \label{eq:SI_nonlocOvlp}
\end{equation}
The proper transformation between the $U^i$ and $W^j$ orbital bases in this case thus includes the metric explicitly through
\begin{equation}
    T^{i\rightarrow j}_{U\rightarrow W} = W^j g^{ij} \left(U^i\right)^\dagger,
    \label{eq:SI_Tnonloc}
\end{equation}
which then enters the exponential transformation operator for many-body objects, $\exp\left[(\vec{c})^\dagger T^{i\rightarrow j}_{U\rightarrow W}\vec{c}\right]$.
Unitary transformation matrices of this form enter the computation of the overlap and Hamiltonian matrices in Eq.~\eqref{eq:EC_HandS}.
In this work, we evaluate this expectation values in the MO basis of the Hamiltonian of Eq.~\eqref{eq:EC_HandS}, which will in general be different than that of $U^i$ and $W^j$.
}

\LK{
The previous considerations including the metric would not be a major issue, if we kept all orbitals in our calculation.
This is however computationally unachievable in general, and instead an active space is chosen.
Then, then presence of $g^{ij}$ can lead to an important problem in the computational chemistry setting: even if the atomic character ($s,p,d,\dots$) of the orbitals in the active space does not change between the $U^i$ and $W^j$ bases, the fact that the AOs are typically localized (eg. as Gaussian orbitals) results in the inner products in Eq.~\eqref{eq:SI_metric} naturally decreasing exponentially when $\boldsymbol{\lambda}_i$ and $\boldsymbol{\lambda}_j$ represent different molecular geometries.
As a consequence, the generalized eigenvalue problem in Eq.~\eqref{eq:EC_HandS} becomes ill-conditioned exponentially quickly along the PES, and one would formally need an exponentially dense grid of sample points to recover the full PES within the EC Ansatz.
}

\LK{
This will happen if we try to perform the EC with the exact orbital matching, which includes the metric $g^{ij}$ in Eq.~\eqref{eq:SI_Tnonloc}.
Instead, we can alleviate the orbital mismatch due to the exponential decrease of the metric within the active space by making the substitution $g^{ij}\rightarrow \sqrt{S_j}\sqrt{S_i}$ in Eq.~\eqref{eq:SI_Tnonloc}.
This corresponds to the orbital matching ignoring the metric described in Sec. IIIB.
Intuitively, this simplification ignoring the spatial displacement of the AOs with the change in orbital geometry, while still accounting for the changes in the MO composition in terms of those AOs.
An even more insensitive approximation would be to assume $\braket{\alpha(\boldsymbol{\lambda}_i)|\beta(\boldsymbol{\lambda}_j)} = \delta_{\alpha \beta}$
, which indeed would correspond to not rotating the MO basis at all between parameter points $\boldsymbol{\lambda}_i$, which is also briefly presented in Sec. IIIB.
In our current implementation in pyscf, we use the function \emph{fci.addons.transform\_ci\_for\_orbital\_rotation} to apply the exponential transformation operator the the FCI ground states for the given choice of rotation matrix $T^{i\rightarrow j}_{U\rightarrow W}$.
In summary, these choices are
\begin{equation}
    T^{i\rightarrow j}_{U\rightarrow W} = \begin{cases}
        \mathbb{I} & \mathrm{No\ Rotation}\\
        W^j \sqrt{S_j}\sqrt{S_i} \left(U^i\right)^\dagger & \mathrm{No\ metric}\\
        W^j g^{ij} \left(U^i\right)^\dagger & \mathrm{Full\ rotation}
    \end{cases}.
    \label{eq:SI_matchings}
\end{equation}
}
\end{document}